\documentstyle[12pt]{article}
\def\be{\begin{equation}}
\def\ee{\end{equation}}

\begin{document}

\title{Atomic Quantum Computer}
\author{
 I.V. Volovich\thanks{Permanent address: Steklov Mathematical
Institute, Gubkin St.8, GSP-1, 117966, Moscow; volovich@mi.ras.ru}\\
{\it Science University of Tokyo}\\
{\it Department of Information Sciences}\\
{\it Noda City, Chiba 278, Japan}\\
{\it  volovich@mo.is.noda.sut.ac.jp}
}
\date{}
\maketitle
\begin {abstract}
The current proposals for the realization of
quantum computer such as NMR, quantum dots and
trapped ions are based on the using of an atom or an ion as one qubit. In
these proposals a quantum computer consists from several atoms and the
coupling between them provides the coupling between qubits necessary for a
quantum gate. We discuss whether a {\it single} atom can be used as a
quantum computer. Internal states of the atom serve to hold the quantum
information and the spin-orbit and spin-spin interaction provides the
coupling between qubits in the atomic quantum computer. In particular one
can use the electron spin resonance (ESR) to process the information encoded in
the hyperfine splitting of atomic energy levels. By using quantum state
engineering one can manipulate the internal states of the natural or
artificial (quantum dot) atom to make
quantum computations.
\end {abstract}
\newpage
\setcounter{equation}{0}


Quantum computers \cite{Ben,Fey,Deu,EJ,Llo} have an information processing
capability much greater than the classical computers. Considerable progress
in quantum computing has been made in recent years. A number of quantum
algorithms have been developed \cite{Sho, Gro} and experimental
implementations of small quantum computers have been achieved \cite{Ger,
Cor, Jon}. In particular such realizations of quantum computers as NMR \cite
{Ger, Cor}, ion traps \cite{CZ}, cavity QED \cite{Tur} and quantum dots \cite
{DiV} have been proposed.

The proposed technologies for realization of quantum computer have serious
intrinsic limitations \cite{Pre}. In particular NMR devices suffer from an
exponential attenuation of signal to noise as the number of qubits increase
and an ion trap computer is limited by the frequencies of the vibrational
modes in the trap. In this note we discuss a possible realization of quantum
computer which perhaps can help to avoid these limitations.

Basic elements of quantum computer are qubits and logic elements (quantum
gates). A qubit is a two-state quantum system with a prescribed
computational basis. The current proposals for the experimental realization
of quantum computer are based on the implementation of the qubit as a
two-state atom or an ion. Quantum computer in these schemes is a {\it 
molecular machine} because it is built up from a number of coupled atoms or
quantum dots. Here we propose to do quantum
computations using a {\it single} atom . In this scheme 
the atomic quantum computer is a single atom.
It is interesting to study such an {\it atomic machine} theoretically but
it could have also some advantages for the practical realization with
respect to the molecular machines.

It is well known that in atomic physics the concept of the individual 
state of an electron in an atom
is accepted and one proceeds from the self-consistent field approximation,
see for example \cite{Sob}.
The state of an atom is determined by the set of the states of the electrons.
Each state of the electron is characterized by a definite value of 
its orbital angular momentum $l$, by the principal 
quantum number $n$ and by the values of the projections of the orbital angular 
momentum $m_l$ and of the spin $m_s$ on the $z$-axis. In the Hartree-Fock
central field approximation
the energy of an atom is completely determined by the 
assignment of the electron
configuration, i.e., by the assignment of the values of 
$n$ and $l$ for all the electrons.

One can implement a single qubit
in atom 
as a one-particle electron state in the self-consistent field approximation and
multi-qubit states as the corresponding multi-particle states represented
by the Slater determinant.

Almost all real spectra can be systematized with respect to 
$LS$ or $jj$ coupling schemes.
Every stationary state of the atom in the $LS$ coupling approximation 
is characterized  by a definite value of the orbital angular momentum
$L$ and the total spin $S$ of the electrons. Under the action of the relativistic 
effects a degenerate level with given $L$ and $S$ is split into a number
of distinct levels (the fine structure of the level), 
which differ in the value of the total angular momentum $J$.
 The relativistic terms in the Hamiltonian
of an atom includes the spin-orbit and spin-spin interaction. There is also the further
splitting of atomic energy levels as a result of the interaction of electrons
with the spin of the nucleus. This is the hyperfine structure of the levels.

 One can use these
interactions to build  quantum logic gates.

As a simple example let us discuss how the hyperfine splitting can be used
to do quantum computations on a single atom.
Let us consider the Hamiltonian which includes both nucleus and 
electron for a case of quenched orbital angular momentum. If one assumes
that the electron spin Zeeman energy is much bigger than the hyperfine coupling
energy then one gets an approximate Hamiltonian \cite{Sli}
$$
{\cal H}=g\beta HS_z - \gamma_n \hbar HI_z+AS_zI_z
$$
Here $S_z$ and $I_z$ are the electron and nuclear spin operators, $H$ is the magnetic
field which is parallel to the $z$-axis, $\beta$ is the Bohr magneton, $\gamma_n$
is the nuclear gyromagnetic ratio, $A$ is the hyperfine 
coupling energy and $g$ is the $g$-factor.

Let us consider the simplest case of nuclear and electron spins of 1/2.
Then a single qubit 
is a nuclear spin $|m_I>$ and electron spin $|m_S>$
function, where $m_I$ and $m_S$ stand for eigenvalues of $I_z$ and $S_z$.
The two-qubit states are the eigenfunctions of the Hamiltonian ${\cal H}$
and they are given by the product of the nuclear spin and electron spin functions
$$
|m_I,m_S>=|m_I> |m_S>
$$
The coupling used to produce magnetic resonances is an alternating
magnetic field applied perpendicular to the static field.
The possible transitions produced by an alternating field are found
by considering  a perturbing term in the Hamiltonian
$$
{\cal H}_m (t) = (\gamma_e \hbar S_x - \gamma_n \hbar I_x) H_x \cos \omega t
$$
Many of the basic principles of nuclear magnetic resonance apply to electron
magnetic resonance (ESR). However
there are some special features of spin echoes that arise for electron
spin resonance which are not encountered in nuclear magnetic resonance.
This is because in many cases the nuclear quantization direction depends
on the electron spin orientation.

It is well known that any quantum algorithm can be implemented 
with one-qubit rotations and two-qubit controlled-NOT gate, 
see \cite{EJ,Llo,OW}.
The implementation of the controlled-NOT gate by using  pulse
sequences is well known in NMR \cite{Ger,Cor,Jon}.  For example
it can be represented
as a network which includes one qubit Hadamard gates and a 
$4\times 4$ matrix which can be implemented as the following pulse sequence
$$
(90^0I_z)(90^0S_z)(-90^02I_zS_z)
$$
Two-qubite  realizations of
the Deusch-Jozsa algorithm and the Grover algorithm have been 
accomplished
using NMR spectroscopy of spin 1/2 nuclei of appropriate molecules
in solution \cite{Ger, Cor, Jon}. One can use the similar technique
in the case of ESR. 

If computers are to become much smaller in the
future, the miniaturization  might lead to the atomic quantum computer.
One of advantages of the atomic quantum computer is that 
quantum state of a single atom can be stable against decoherence,
for a discussion of the decoherence problem in quantum computing see 
\cite{Llo, Pre, Vol} and references therein.
Recent experimental and theoretical advances on  quantum state engineering
with a  natural and artificial (quantum dots) atoms 
\cite{Oos, Ye, Enk, Roo, SC} and the 
development of methods
for completely determining the quantum  state of an atom \cite{Var}
show that quantum computations with a single atom should be possible.

To summarize, I propose using a single atom to do quantum computations.
 Such an atom can be also used, of course,
as a part of a computational network.  I discussed the simple realization
of the two-qubit atomic quantum computer based on ESR and hyperfine splitting.
However the idea of atomic quantum computer is more general.  To build a 
multi-qubit atomic quantum computer one has to
use the fine and hyperfine splitting of energy levels to process the
information encoded in the multielectron states.
In principle one can build an atomic quantum computer based on a natural
or artificial (quantum dot) atom .

I am grateful to M. Ohya and N. Watanabe for stimulating discussions
on quantum computing.

\end{document}